\documentclass[conference]{IEEEtran}
\IEEEoverridecommandlockouts
\usepackage{cite}
\usepackage{amsmath,amssymb,amsfonts}
\usepackage{algorithmic}
\usepackage{float} 
\usepackage{svg}
\usepackage{graphicx}
\usepackage{booktabs}
\usepackage{makecell}
\usepackage{textcomp}
\usepackage{multirow}
\usepackage{url}
\usepackage{xcolor}
\usepackage[colorlinks=true,linkcolor=red,citecolor=blue,urlcolor=blue]{hyperref}
\def\BibTeX{{\rm B\kern-.05em{\sc i\kern-.025em b}\kern-.08em
    T\kern-.1667em\lower.7ex\hbox{E}\kern-.125emX}}
\begin{document}

\title{
TourSynbio: A Multi-Modal Large Model and Agent Framework to Bridge Text and Protein Sequences for Protein Engineering
}


\author{
\IEEEauthorblockN{
Yiqing Shen\textsuperscript{1,2,$\dagger$}, Zan Chen\textsuperscript{1,$\dagger$}, Michail Mamalakis\textsuperscript{3,$\dagger$}, Yungeng Liu\textsuperscript{1,6,$\dagger$,$\S$}, \\ 
Tianbin Li\textsuperscript{4}, Yanzhou Su\textsuperscript{4}, Junjun He\textsuperscript{4}, 
Pietro Li\`{o}\textsuperscript{3},
Yu Guang Wang\textsuperscript{1,4,5,*}
}
\IEEEauthorblockA{
\textsuperscript{1}\textit{Toursun Synbio}, Shanghai, China\\
\textsuperscript{2}\textit{Department of Computer Science}, \textit{Johns Hopkins University}, Baltimore, USA\\
\textsuperscript{3}\textit{Department of Computer Science and Technology}, \textit{University of Cambridge}, Cambridge, UK\\
\textsuperscript{4}\textit{Shanghai AI Laboratory}, Shanghai, China\\
\textsuperscript{5}\textit{Institute of Natural Sciences}, \textit{Shanghai Jiao Tong University}, Shanghai, China\\
\textsuperscript{6}\textit{Department of Computer Science}, \textit{City University of Hong Kong}, Hong Kong, China\\
{\footnotesize\textsuperscript{$\dagger$}Equal contribution. 
\textsuperscript{$\S$}Work done during the internship at Toursun Synbio. 
\textsuperscript{*}Corresponding author.}\\
{\footnotesize yshen92@jhu.edu  \qquad yuguang.wang@sjtu.edu.cn}
}}

\maketitle

\begin{abstract}
The structural similarities between protein sequences and natural languages have led to parallel advancements in deep learning across both domains.
While large language models (LLMs) have achieved much progress in the domain of natural language processing, their potential in protein engineering remains largely unexplored. 
Previous approaches have equipped LLMs with protein understanding capabilities by incorporating external protein encoders, but this fails to fully leverage the inherent similarities between protein sequences and natural languages, resulting in sub-optimal performance and increased model complexity.
To address this gap, we present TourSynbio-7B, the first multi-modal large model specifically designed for protein engineering tasks without external protein encoders. 
TourSynbio-7B demonstrates that LLMs can inherently learn to understand proteins as language.
The model is post-trained and instruction fine-tuned on InternLM2-7B using ProteinLMDataset, a dataset comprising 17.46 billion tokens of text and protein sequence for self-supervised pretraining and 893K instructions for supervised fine-tuning.
TourSynbio-7B outperforms GPT-4 on the ProteinLMBench, a benchmark of 944 manually verified multiple-choice questions, with 62.18\% accuracy. 
Leveraging TourSynbio-7B's enhanced protein sequence understanding capability, we introduce TourSynbio-Agent, an innovative framework capable of performing various protein engineering tasks, including mutation analysis, inverse folding, protein folding, and visualization. 
TourSynbio-Agent integrates previously disconnected deep learning models in the protein engineering domain, offering a unified conversational user interface for improved usability.
Finally, we demonstrate the efficacy of TourSynbio-7B and TourSynbio-Agent through two wet lab case studies on vanilla key enzyme modification and steroid compound catalysis. 
Our results show that this combination facilitates protein engineering tasks in wet labs, leading to higher positive rates, improved mutations, shorter delivery times, and increased automation.
The model weights are available at \url{https://huggingface.co/tsynbio/Toursynbio} and codes at \url{https://github.com/tsynbio/TourSynbio}.
\end{abstract}

\begin{IEEEkeywords}
Deep Learning, Multi-modal Large Model, Protein Engineering, AI Agent.
\end{IEEEkeywords}

\section{Introduction}

\begin{figure}[htbp!]
\centering\centerline{\includegraphics[width=0.85\linewidth]{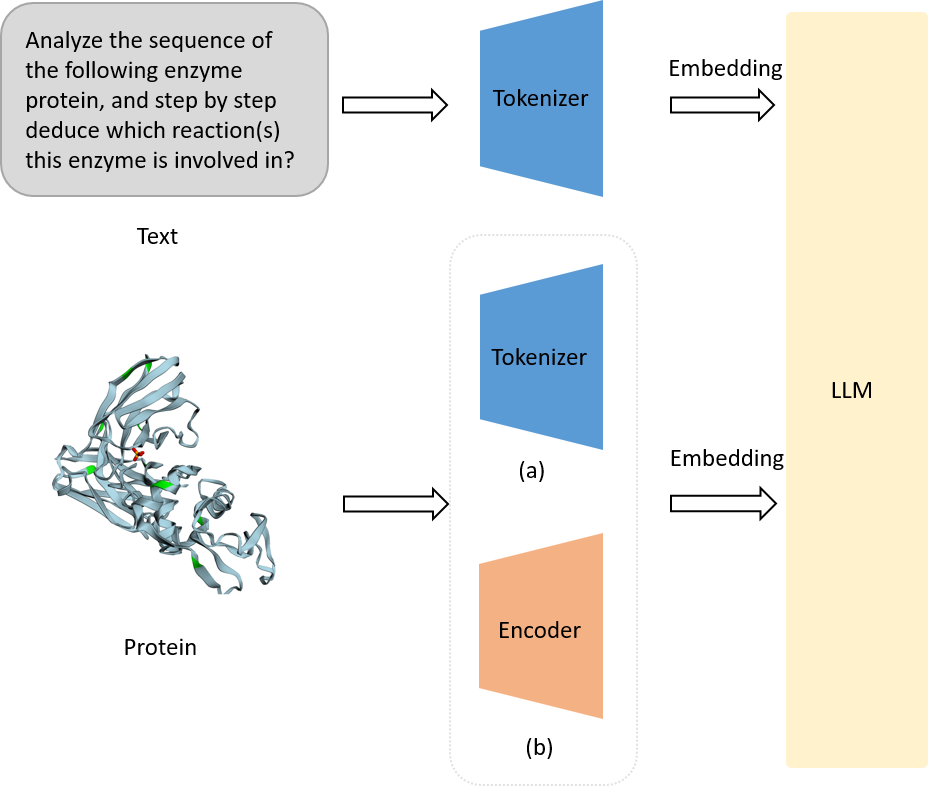}
}
\caption{Illustration of the traditional LLM and TourSynbio-7B for protein sequence understanding. (a) Traditional methods use external encoders to process protein sequences before feeding them into LLMs. (b) The proposed TourSynbio-7B method directly processes protein sequences using an LLM without the need for external encoders, thus simplifying the workflow and improving efficiency.}
\label{fig:intro}
\end{figure}

Protein engineering enables the modification and optimization of protein sequences or structures for diverse applications, revolutionizing our ability to manipulate biological systems at the molecular level \cite{leisola2007protein,listov2024opportunities}.
Deep learning approaches offer higher efficiency and better performance in protein engineering tasks by efficiently processing vast amounts of protein data, capturing complex patterns in sequences, predicting protein properties and structures with increasing accuracy, and facilitating rapid exploration of optimal protein designs \cite{defresne2021protein,dauparas2022robust,wang2018computational}. 
In protein engineering, protein sequences serve as the fundamental data format, often referred to as the ``language of life sciences'' due to their role in encoding biological information \cite{bairoch2000swiss,heinzinger2019modeling}.
The inherent sequential similarities between protein sequences and natural language have already led to the parallel development of foundation models, namely protein language models \cite{rives2019biological} and large language models (LLMs) \cite{gpt4}.
Because LLMs have demonstrated strong capabilities in text understanding \cite{18}, recent research has begun exploring their potential in protein understanding through multi-modal large models \cite{lv2024prollama,wang2024protchatgpt}.
Previous attempts have focused on integrating protein sequences or structures in the form of protein graphs with textual content using extra encoders \cite{xu_protst_2023, zhou_protein_2023}, as depicted in Fig.~\ref{fig:intro}(a). 
However, these approaches fail to fully leverage the intricate connections between protein sequences and natural language, leading to higher model complexity and sub-optimal performance.
This limitation highlights the need for a model that can directly understand and process protein sequences without relying on external encoders, potentially improving both efficiency and performance in protein engineering tasks. 
To address this gap, we present TourSynbio-7B, the first multi-modal large model specifically designed for protein engineering tasks without external protein encoders as shown in Fig.~\ref{fig:intro}(b).
TourSynbio-7B is post-trained and instruction fine-tuned on InternLM2-7B \cite{in} using ProteinLMDataset \cite{data}, demonstrating that LLMs themselves can learn to understand proteins in the form of language.

However, the conversation capabilities in multi-modal large models like TourSynbio-7B are not sufficient for protein engineering tasks, as they need to address complex challenges like protein structure prediction \cite{11}, evolution analysis, and \textit{de novo} protein design \cite{6,13}.
AI agent frameworks offer a potential solution by allowing LLMs to address domain-specific tasks through relevant tools \cite{4}.
However, current implementations of AI agents typically limit themselves to common APIs and simple AI models, such as text-to-image generation \cite{kapoor2024ai}. 
Protein engineering involves more complex tasks and APIs, arising from the need to integrate various specialized tools and models, such as protein folding models \cite{d3}, molecular dynamics simulations \cite{hollingsworth2018molecular}, structure-function prediction tools \cite{s7,s6} and \textit{etc} \cite{s8,outeiral2024codon} that takes protein sequence as input. 
This highlights the need for AI agents that can effectively integrate complex, domain-specific protein engineering APIs or models.

Previous agent works like ProteinEngine attempt to build an AI agent framework for protein engineering based on GPT-3.5 by assigning three roles to LLMs, namely task delegation, specialized task resolution, and effective communication of results \cite{prev}. 
However, GPT-3.5 lacks specialized training in protein sequences, hence it struggles to accurately interpret protein motifs, predict mutation impacts, and understand sequence-structure relationships in the agent.
Additionally, ProteinEngine requires calling the LLM multiple times, which reduces efficiency.
To address these limitations, we propose TourSynbio-Agent, an innovative framework built upon TourSynbio-7B's enhanced protein sequence understanding capabilities. 
TourSynbio-Agent integrates previously disconnected deep learning models in the protein engineering domain, offering a unified conversational user interface for improved usability while preserving efficiency with a keyword-matching scheme for calling different models.

The major contributions of this paper are four-fold. 
Firstly, we present TourSynbio-7B, a novel multi-modal large model that bridges the protein sequence and text understanding. 
It demonstrates LLMs can inherently understand proteins as a new language without relying on external protein encoders. 
Second, we present how the user can utilize TourSynbio-7B as an innovative agent framework to perform complex protein engineering tasks, including mutation analysis, inverse folding, and visualization. 
Third, we present a human-centered interface that combines TourSynbio-7B and TourSynbio-Agent, providing researchers and engineers with a unified conversational user interface. 
This interface simplifies the execution of complex protein engineering workflows, enhancing usability and accessibility for both experts and non-experts in the field. 
Finally, we demonstrate the practical efficacy of our approach through two wet lab case studies, showcasing how the combination of TourSynbio-7B and TourSynbio-Agent facilitates real-world protein engineering tasks, leading to higher positive rates, improved mutations, shorter delivery times, and increased automation in wet lab settings.

\begin{table*}[htbp]
\centering
\caption{
Overview of the diverse functionalities integrated into TourSynbio-Agent.
It showcases the range of specialized models and tools incorporated into our system, covering critical aspects of protein engineering such as mutation prediction, protein folding and inverse folding, protein design and reconstitution, and auxiliary tools for sequence analysis and visualization. 
Each model is described with its specific function, input requirements, and output format, illustrating the breadth and depth of capabilities offered by TourSynbio-Agent in addressing complex protein engineering tasks.
}\label{tab:agent}
\begin{tabular}{@{}p{2.2cm}p{1.8cm}p{4cm}p{4cm}p{4cm}@{}}
\toprule
Functionality & Model & Description & Input & Output \\
\midrule
\multirow{5}{*}{Mutation Prediction} & \multirow{2}{*}{ESM-1v \cite{s6}} & Zero-shot prediction of mutation effects on protein function & Protein sequence, mutation offset, and additional parameters & \texttt{CSV} file with detailed mutation effects \\
& \multirow{2}{*}{SaProt \cite{s7}} & Protein language model integrating sequence and structure information & Sequence and structure information in a letter string format & Quantitative mutation score and mutation dictionary \\
\midrule
\multirow{4}{*}{\makecell[l]{Protein Folding \\ and Inverse Folding}}& \multirow{2}{*}{ESMFold \cite{d3}}  & Structure prediction using masked language modeling without multiple sequence alignment  & Raw protein sequence  & Residue-level pLDDT and \texttt{PDB} file of predicted structure \\
& \multirow{2}{*}{ESM-IF1 \cite{s5}} & Transformer-based model for single-site mutation prediction & Original protein sequence  &  Novel protein sequence with predicted mutations \\
& GraDE-IF \cite{yi2024graph} & Graph-based model for inverse folding & Protein sequence & De novo protein sequence \\
\midrule
\multirow{8}{*}{\makecell[l]{Protein Design \\and Reconstitution}} & \multirow{3}{*}{AntiFold \cite{s4}} & Model for designing sequences fitting antibody variable domain structures & PDB code, heavy chain sequence, and additional parameters & \texttt{CSV} file with design metrics and \texttt{FASTA} file of designed sequences \\
& \multirow{2}{*}{Chroma \cite{s3}} & Generative model for programmatic protein design & Desired chain length and design constraints & \texttt{PDB} file of the generate protein \\
& \multirow{3}{*}{LigandMPNN \cite{s1}} & Enhanced version of ProteinMPNN for protein reconstruction considering ligand interactions & PDB code, temperature parameter, and additional constraints & Reconstructed \texttt{PDB} file with optimized protein structure  \\
\midrule
\multirow{2}{*}{Codon Analysis} & \multirow{2}{*}{CaLM \cite{outeiral2024codon}} & Generate high-dimensional embeddings for biological sequences & DNA or protein sequence & Numerical embedding vector representing sequence features \\
\midrule
 \multirow{2}{*}{Visualization} & \multirow{2}{*}{PyMOL} & 3D visualization tool for protein molecules & PDB code or file & Interactive 3D visualization of the protein structure \\
\bottomrule
\end{tabular}
\end{table*}

\section{Methods}

\subsection{TourSynbio-7B}
TourSynbio-7B is built upon the InternLM2-7B \cite{in}, leveraging its language understanding capabilities for both English and Chinese as a foundation for protein sequence comprehension. 
InternLM2-7B utilizes a transformer-based architecture with 32 layers, 4096 hidden dimensions, and 8 attention heads \cite{in}.
For post-training and instruction fine-tuning, we utilized the ProteinLMDataset \cite{data}, a dataset specifically designed for protein sequence understanding. 
Specifically, it comprises 17.46 billion tokens for self-supervised learning and 893,000 instructions for supervised fine-tuning. 
The self-supervised portion is strategically categorized into three segments, namely Chinese-English text pairs in protein science (0.69\%), protein sequence-English text pairs (41.51\%), and protein-related English text (57.80\%). 
This diverse composition of ProteinLMDataset enables TourSynbio-7B to develop a robust understanding of protein-related concepts across multiple languages and data formats.
The post-training process involved two main phases, namely the initial training on 4K context texts, followed by a transition to high-quality 32K texts for further training, which allows TourSynbio-7B to effectively capture long-term dependencies in protein sequences and related scientific literature. 
The instruction fine-tuning phase utilized the 893,000 instructions from ProteinLMDataset, covering various protein engineering tasks and concepts, to enhance the TourSynbio-7B's ability to follow complex protein-related instructions and perform specific protein engineering tasks.
All the training was conducted using the \texttt{XTuner} framework \footnote{\url{https://github.com/InternLM/xtuner}}, which enables efficient scaling across multiple GPUs. 
Through this training process, TourSynbio-7B has developed the capability to understand and process protein sequences as a form of language, without the need for external protein encoders.

\begin{figure}[htbp!]
\centering\centerline{\includegraphics[trim={0.0cm 0.cm 0.05cm 0.05cm},clip,width=\linewidth]{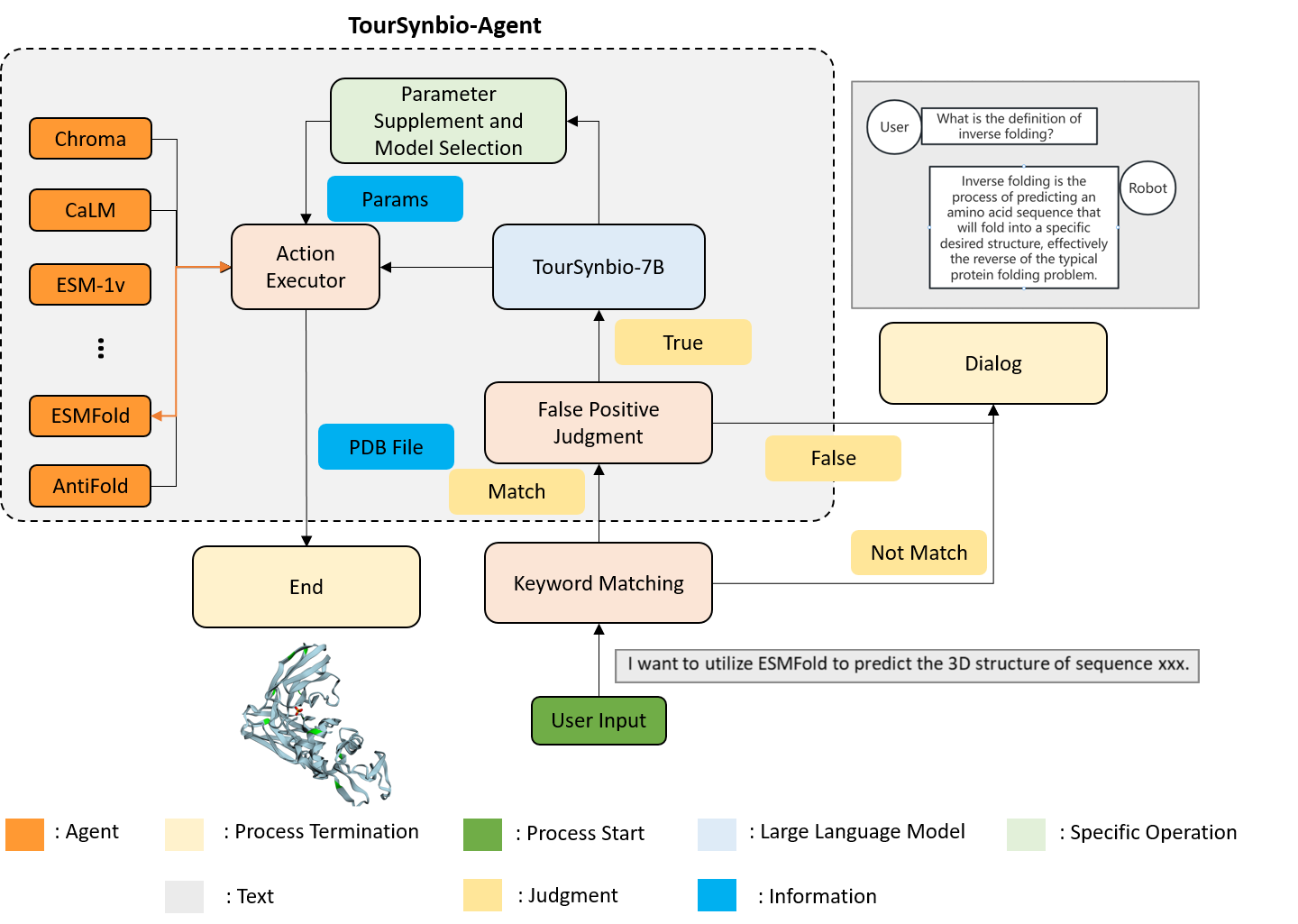}}
\caption{The overall workflow of TourSynbio-Agent from initial user input to final task execution.
It encompasses several stages: (1) Intent classification using TourSynbio-7B to determine if an agent call is required; 
(2) Keyword-based agent selection with fuzzy matching to identify the appropriate agent; 
(3) User-guided selection for functionally similar agents when necessary; 
(4) Parameter extraction and interactive validation to ensure accurate inputs; and (5) Agent execution. 
The process incorporates multiple decision points and feedback loops, allowing for efficient handling of various scenarios, including false positives and ambiguous user inputs. 
This design ensures robust and accurate execution of protein engineering tasks while maintaining a user-friendly interface.
}
\label{ex3}
\end{figure}

\subsection{TourSynbio-Agent}
Building upon TourSynbio-7B, we elaborate on TourSynbio-Agent, the AI agent framework designed to execute complex protein engineering models, as illustrated in Fig.~\ref{ex3}.
TourSynbio-Agent is built on the Lagent framework \cite{lagent2023}.

\subsubsection{Intent Classification}
In this stage, TourSynbio-7B analyzes the user's input to determine whether a specific agent needs to be called. 
When certain words or phrases like ``why'' or ``introduce'' typically indicate that the user is seeking general information rather than requesting an agent-specific task.
These inputs are directed to the standard dialogue process.
For inputs lacking clear keywords, TourSynbio-7B utilizes a custom-designed prompt to classify the user's intent.
%
%
%

\subsubsection{Keyword-Based Agent Selection with Fuzzy Matching}
Given the integration of 10 different functional agents in TourSynbio-7B as depicted in Table~\ref{tab:agent}, relying solely on the language model to determine agent calls could lead to inaccuracies. 
To enhance the accuracy of agent call request recognition, we implemented a keyword-matching mechanism, which utilizes independent keyword tables for each agent, containing specific terms relevant to their function.
%
For instance, the inverse folding agent's table includes keywords such as ``inverse folding'' and ``reconstruction''.
When TourSynbio-7B receives user input, it systematically traverses the keyword table of each agent, checking for matches.
If corresponding keywords are identified, the system activates the appropriate agent for processing. 
If no match is found, it proceeds to check the next agent's keyword table.
This design of independent keyword tables effectively prevents erroneous agent activation.
To further improve the accuracy and robustness of the agent workflow and enhance user experience, we incorporated fuzzy matching capabilities, which accommodate potential spelling errors by users.
It ensures that even if a keyword is input incorrectly, the system can still recognize and call the corresponding agent accurately. 
For example, minor misspellings like ``inverse foding'' or ``recontruction'' would still trigger the inverse folding agent.

\begin{center}
\fcolorbox{black}{gray!10}{\parbox{0.9\linewidth}{
\textbf{TourSynbio-Agent Intent Classification Prompt:}\\
You are an information extraction assistant. You need to determine whether the user's input intends to call any agent and return \textit{True} or \textit{False}.\\
It should be noted that when a user inputs a sequence of letters or numbers in text, we tend to assume that the user is entering the parameters of the agent.\\
But you need to note that model names do not belong to the category of alphabetical sequences, such as Chroma, CaLM, SaProt, ESM-IF1. When these sequences appear, you need to carefully determine whether the user is in a normal conversation or wants to call the agent.\\

The example is as follows:\\
Q: Reverse fold PDB file 5YH2 with chain ID A, temperature set to 1.0, and sampling quantity of 1.\\
A: \{\textit{True}\}\\
Q: What is inverse folding?\\
A: \{\textit{False}\}\\

Please extract information from the following sentence:
Q: user\_input
}}
\end{center}

\subsubsection{User-Guided Selection for Functionally Similar Agents}
TourSynbio-Agent incorporates an optional user selection function to address scenarios where multiple agents possess similar functionalities.
For example, in the case of mutation prediction, where both SaProt \cite{s7} and ESM-1v \cite{s6} agents offer highly correlated capabilities.
In such instances, conventional keyword matching might not accurately direct users to the most suitable agent. To address this, we have implemented a selection interface for agents with overlapping functions.
When the system identifies a specific keyword (such as ``mutation prediction'') that could potentially trigger multiple agents, it prompts the user to manually select the desired agent.
This interface presents users with a choice between the available options, such as SaProt and ESM-1v for mutation prediction tasks, where users can thus make informed decisions based on their specific requirements.
This user-driven selection process not only mitigates the limitations of keyword-based routing but also empowers users with greater control over their protein engineering workflows.

\subsubsection{Parameter Extraction and Interactive Validation}
After successful keyword matching, TourSynbio-7B employs a validation process to confirm the user's intended agent call, thereby mitigating false positives that may arise during the keyword-matching phase. 
This stage operates on the assumption that users are cognizant of their desired agent and its associated parameter requirements.
To facilitate accurate parameter extraction, we have developed tailored information extraction methods for each agent using prompt engineering by leveraging TourSynbio-7B's natural language processing capabilities to parse user input effectively. 
For instance, consider the following prompt designed for the ESMFold \cite{d3} agent:

\begin{center}
\fcolorbox{black}{gray!10}{\parbox{.9\linewidth}{
\textbf{ESMFold Agent Parameter Extraction Prompt:}\\
You are an information extraction assistant. Please extract the protein sequence from the following sentence and return it in the format of `\{sequence: xxx\}'.\\

Example:\\
Q: Perform structure prediction on the sequence QERLKSIVRILERSKEPVSG.\\
A: \{sequence: QERLKSIVRILERSKEPVSG\}\\

Please extract information from the following sentence:
Q: user\_input
}}
\end{center}

This prompt utilizes a few-shot prompting approach to enhance TourSynbio-7B's ability to parse information accurately. 
It returns results in \texttt{JSON} format, facilitating efficient information extraction and subsequent processing.
Upon receiving user input, TourSynbio-7B attempts to extract all required parameters. 
If any essential parameters are missing or ambiguous, TourSynbio-7B generates an interactive interface. 
This interface allows users to verify the accuracy of extracted parameters, ensuring that the agent operates with the correct inputs. 
It also provides an opportunity for users to supply any optional parameters that may enhance the agent's performance or customize its output.

%

\subsubsection{Agent Execution}
After completing all the previous steps, our TourSynbio-7B calls the corresponding agent to execute the task and return the result. 
If keyword matching fails to identify a suitable agent but the user's intention is clear through text analysis, TourSynbio-7B automatically determines which agent to call and parses the required parameters to complete the task. 
%
%

\begin{figure*}[htbp!]
\centering\centerline{\includegraphics[width=\linewidth]{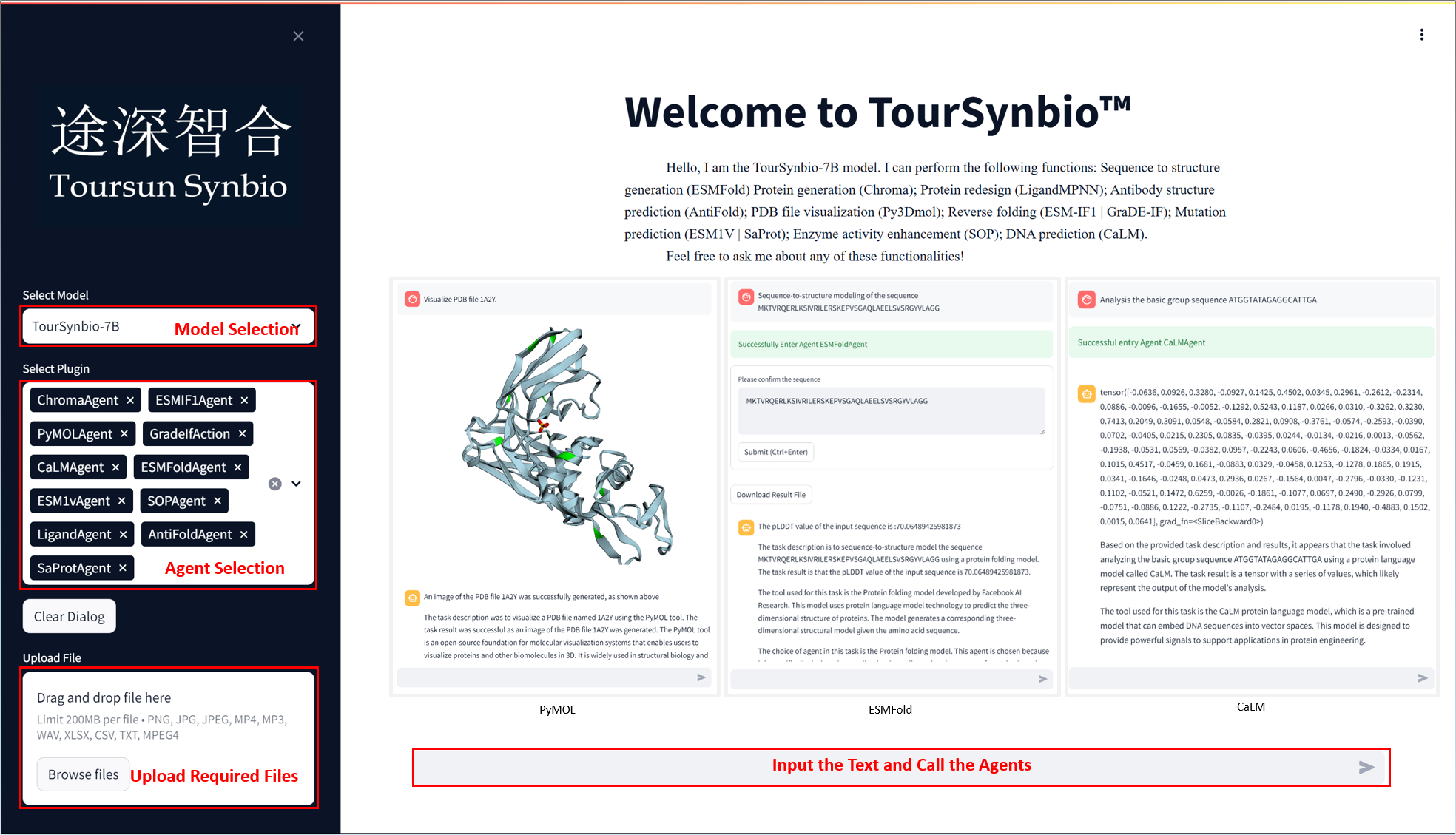}}
\caption{
The Human-centered conversational user interface of TourSynbio-Agent. 
It integrates four key components: (1) Model Selection, allowing users to choose between TourSynbio-7B and other language models; 
(2) Agent Selection, enabling customization of active protein engineering tools; 
(3) File Upload, supporting various file formats for data input; and (4) Text Input area for natural language interactions. 
The interface also showcases real-time visualizations and results from various protein engineering tasks, including the PyMOL, ESMFold \cite{d3}, and CaLM \cite{outeiral2024codon} agents.
This intuitive design facilitates seamless interaction between users and the AI-driven protein engineering models, enhancing workflow efficiency and accessibility for both experts and non-experts in the field.
A video demo is available at \href{https://github.com/tsynbio/TourSynbio/blob/main/demo/video_demo.mp4}{here on GitHub}.
%
}
    \label{ex4}
\end{figure*}

\subsection{Human-centered Conversational User Interface}
Our human-centered conversational user interface further enhances the user experience and system usability of TourSynbio-7B and TourSynbio-Agent which provides an intuitive and user-friendly interface, as illustrated in Fig.~\ref{ex4}. 
The interface design is based on the following key principles:
\begin{itemize}
\item \textbf{User-friendliness.}
The interface adheres to principles of simplicity and intuitiveness, enabling effortless user interaction. 
Leveraging TourSynbio-7B to process nature language and protein sequence, users can input requests using everyday language without the need to learn complex command syntax of protein engineering. 
It can reduce the learning curve, making the interaction with protein engineering models accessible to users with diverse backgrounds and expertise levels.
\item \textbf{Customizability.}
It offers flexible interface options, allowing users to tailor the layout and functional modules to their specific needs and preferences. 
Users can customize the workflow by selecting specific agents, adjusting the interface theme, and modifying color schemes to enhance their comfort and productivity. 
This personalization capability enables users to create an optimal working environment that aligns with their individual work habits and preferences.
\end{itemize}

Our CUI includes three functional modules.
\subsubsection{Model Selection} 
Our CUI offers users the flexibility to choose from a range of large language models, including TourSynbio-7B and GPT-4-turbo, to drive the agent framework. 
This feature provides several key advantages in the context of protein engineering tasks. 
Specifically, users can select the most appropriate model for different tasks, ensuring optimal processing and performance. 
For instance, TourSynbio-7B may be preferred for tasks requiring specialized protein sequence understanding, while GPT-4-turbo might be chosen for more general language tasks. 
The interface is designed with extensibility in mind, accommodating the addition of other LLMs in the future.

\subsubsection{Agent Selection}
Our CUI also allows users to customize the set of agents activated for each task, enhancing the professionalism and accuracy of task execution.
By selecting specific agents, users can ensure that the most suitable tools and functions are employed for their current protein engineering project. 
This flexibility enables users to fine-tune the system's behavior according to different scenarios, avoiding potential errors that might occur with automatic selection. 
Users can activate or deactivate multiple agents based on task requirements, optimizing the processing of each protein engineering task while enhancing their sense of control and trust in the system.

\subsubsection{File Upload}
The interactive interface supports file uploads to facilitate various agent tasks. 
Users can easily upload files through drag-and-drop or browsing functions, with support for multiple file types including \texttt{PDB}, \texttt{FASTA}, \texttt{CSV}, \texttt{TXT}. 
This feature enables users to seamlessly incorporate necessary data when handling complex protein engineering tasks, thereby improving work efficiency and accuracy in task completion.

\section{Experiments}

\subsection{Implementation Details}
We used consistent hyperparameters across both phases. The learning rate was set to 0.0002, with a batch size of 16. We employed the AdamW optimizer and alternated between LinearLR and CosineAnnealingLR for parameter scheduling. 
The training method utilized LoRA with an $\alpha$ value of 16 and a dropout rate of 0.1. 
We trained the TourSynbio-7B on eight NVIDIA A100 GPUs, each with 80GB of memory.

\subsection{Quantitative Evaluations}

\subsubsection{ProteinLMBench and Evaluation Metric}
To assess the performance of TourSynbio-7B and compare it with other LLMs, we utilized ProteinLMBench \cite{data}.
%
ProteinLMBench consists of 944 manually verified multiple-choice questions, covering a wide range of protein-related topics including protein general knowledge, literature understanding, property prediction, function prediction, and protein design \cite{data}.
The evaluation metric for our experiments was accuracy, defined as the percentage of correctly answered questions on ProteinLMBench.

\subsubsection{Experimental Setups}
We compare with the TourSynbio-7B with various state-of-the-art LLMs in a zero-shot manner, including Falcon-7b \cite{f}, Qwen1.5-7B \cite{q}, Moonshot \cite{moon}, Mistral-7B-Instruct-v0.2 \cite{m}, Baichuan2-7B-Chat \cite{b}, LLaMA-2-7B-Chat-hf \cite{ll}, InternLM-Chat-20B, InternLM2-Chat-7B \cite{in}, ChatGLM3-6B \cite{ch}, Yi-6B-Chat \cite{yi}, GPT3.5-turbo, and GPT4-turbo \cite{gpt4}. 
For the ablation study, we compared TourSynbio-7B trained with instruction fine-tuning only (denoted as `w/o PT') and the one with full settings, \textit{i}.\textit{e}., trained with both instruction fine-tuning and post-training.

\begin{table}[htbp]
  \caption{Performance comparison of TourSynbio-7B against state-of-the-art LLMs on ProteinLMBench. Correct Rate indicates accuracy, and Inference Time is measured in minutes. 
  TourSynbio-7B (w/o PT) refers to the model without post-training. 
  \textbf{Bold} indicates the best performance.
  %
  }
  \label{tab:comparison}
  \centering
  \setlength{\tabcolsep}{1mm}{
  \begin{tabular}{lcc}
    \toprule
    \multicolumn{3}{c}{Protein models score}                   \\
    \cmidrule(r){1-3}
    Model & Correct Rate (\%) & Inference Time (min)\\
    \midrule
    GPT4-turbo \cite{gpt4}                & 57.94 & 15.52\\
    InternLM2-20B \cite{in}               & 57.52 & 47.20\\    
    GPT3.5-turbo \cite{gpt4}                         & 55.19 & 21.03\\
    InternLM2-7B \cite{in}                & 54.98 & 19.23\\
    InternLM2-Chat-7B \cite{in}           & 54.76 & 35.58\\
    InternLM2-Chat-20B \cite{in}          & 51.38 & 31.11\\
    Yi-6B \cite{yi}                       & 50.85 & 59.05\\
    Mistral-7B-Instruct-v0.2 \cite{m}     & 50.11 & 13.00\\
    ChatGLM3-6B \cite{ch}                 & 48.94 & 8.00\\
    Baichuan2-7B \cite{b}                                                 & 44.49 & 16.37\\
    InternLM-Chat-20B \cite{in}           & 40.54 & 66.00\\   
    LLaMA2-7B \cite{ll}                   & 39.64 & 64.00\\ 
    Moonshot \cite{moon}                  & 38.26 & 16.25\\
    Qwen1.5-7B \cite{q}                   & 21.73 & 13.00\\
    Falcon-7B-Instruct \cite{f}           & 20.55 & 25.42\\
    Falcon-7B \cite{f}                    & 19.17 & 15.55\\ 
    \midrule
    TourSynbio-7B        & \pmb{62.18} &22.34\\
    TourSynbio-7B (w/o PT) & 58.26 & 21.36\\
    \bottomrule
  \end{tabular}}
\end{table}

\subsubsection{Results}

Table \ref{tab:comparison} presents the performance of LLMs on the ProteinLMBench, comparing their accuracy scores and inference times. 
TourSynbio-7B demonstrates superior performance, achieving the highest accuracy of 62.18\% among all models evaluated, including larger models and commercial APIs.
Notably, TourSynbio-7B outperforms GPT4-turbo, by a margin of 4.24 percentage points (62.18\% vs. 57.94\%). 
This result is impressive given that GPT4-turbo is a much larger model with more extensive training data.
The ablation study results show the effectiveness of our two-stage training approach. TourSynbio-7B with full settings (both post-training and instruction fine-tuning) achieves a 3.92 percentage point improvement over the version without post-training (62.18\% vs. 58.26\%). 
It underscores the importance of the post-training self-supervised learning phase in enhancing the model's protein understanding capabilities.
Among the 7B-parameter models, TourSynbio-7B significantly outperforms its counterparts. 
For instance, it surpasses InternLM2-7B by 7.2 percentage points (62.18\% vs. 54.98\%) and Mistral-7B-Instruct-v0.2 by 12.07 percentage points (62.18\% vs. 50.11\%). 
This demonstrates the effectiveness of our specialized training approach for protein-related tasks.
Regarding inference time, TourSynbio-7B maintains competitive efficiency. 
Its inference time of 22.34 minutes on ProteinLMBench is comparable to or better than many other models, including some with lower accuracy scores. 
This indicates a good balance between performance and computational efficiency.
It's worth noting that TourSynbio-7B even outperforms larger models like InternLM2-20B (57.52\%) and InternLM2-Chat-20B (51.38\%), despite having fewer parameters. 
This suggests that our targeted training on protein-related data is more effective for protein understanding tasks than simply increasing model size.

\subsection{Qualitative Comparison via Case Study}
\begin{figure}[h!]
    \centering
    \centerline{\includegraphics[width=0.65\linewidth]{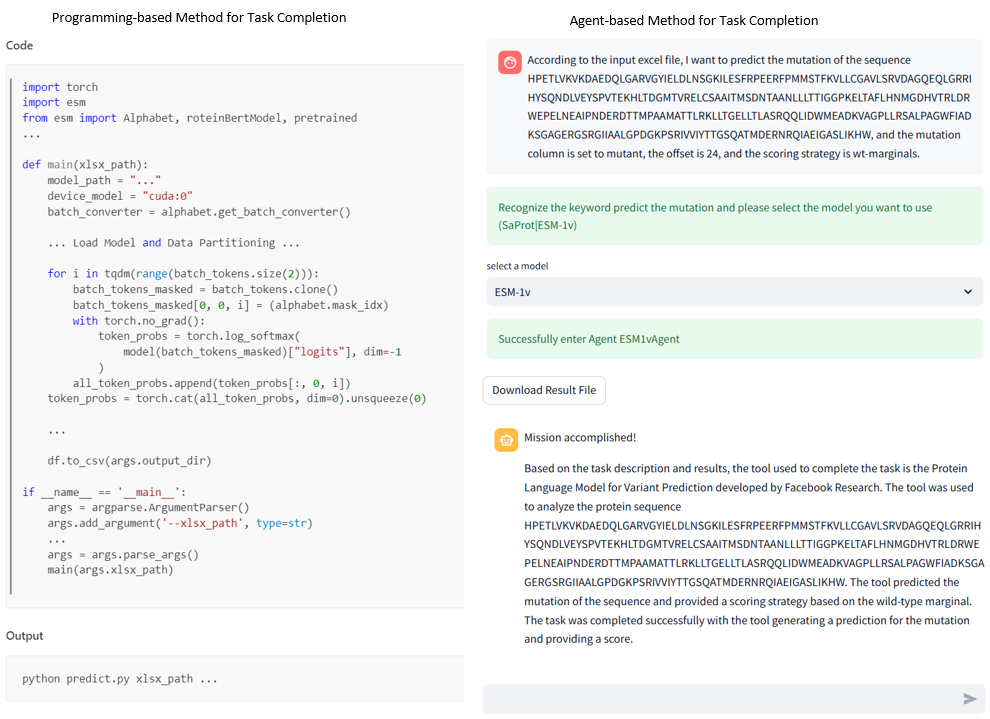}}
    \caption{Comparison between traditional programming-based and TourSynbio-Agent for protein engineering tasks. 
    The left panel illustrates a code snippet for mutation prediction using a deep learning model, while the right panel showcases the TourSynbio-Agent's user-friendly conversational interface for the same task, utilizing the ESM-1v model. 
    This highlights the ease and efficiency of the TourSynbio-Agent, which abstracts complex code into intuitive interactions, thereby streamlining the protein engineering workflow.}
    \label{ex1}
\end{figure}

To illustrate the practical benefits of TourSynbio-Agent over traditional programming methods in protein engineering tasks, we conducted a case study comparing these two approaches for analyzing turnover numbers.
Fig.~\ref{ex1} highlights this comparison, showcasing the key differences in user experience and workflow efficiency.
The left panel of Fig.~\ref{ex1} depicts a conventional programming-based method, requiring users to write and execute Python code for data analysis. 
This method necessitates extensive programming expertise and familiarity with specific libraries such as PyTorch. 
Users must manually handle data loading, model initialization, and result processing, which can be time-consuming and error-prone, particularly for researchers without a strong programming background.
In contrast, the right panel of Fig.~\ref{ex1} demonstrates TourSynbio-Agent's approach, offering a more intuitive and user-friendly interface. 
Users interact with the system using natural language, exemplified by the query, ``I want to predict the mutation of the sequence''.
It abstracts the complexities of code implementation, enabling researchers to concentrate on their scientific inquiries rather than the technical details of programming.
TourSynbio-Agent's conversational interface democratizes access to advanced protein engineering tools, making them accessible to a wider range of researchers, including those lacking extensive programming skills. 
%
%
Furthermore, TourSynbio-Agent adapts to various tasks without requiring users to write or modify code, enhancing research efficiency and reducing the likelihood of errors.
While not explicitly shown in the figure, the interface likely includes options for downloading and visualizing results, further increasing its utility. 
This case study demonstrates how TourSynbio-Agent's human-centered design significantly lowers the barrier to entry for complex protein engineering tasks, making sophisticated analyses more accessible and efficient.

\subsection{Wet Lab Downstream Case Studies}
Finally, to demonstrate the practical applicability of TourSynbio-7B and TourSynbio-Agent, we conducted two downstream case studies, namely vanilla key enzyme modification and steroid compound catalysis. 
These studies aim to showcase how our model can address real-world challenges.

\subsubsection{Vanilla Key Enzyme Modification}
\begin{figure}[h!]
    \centering
    \centerline{\includegraphics[width=\linewidth]{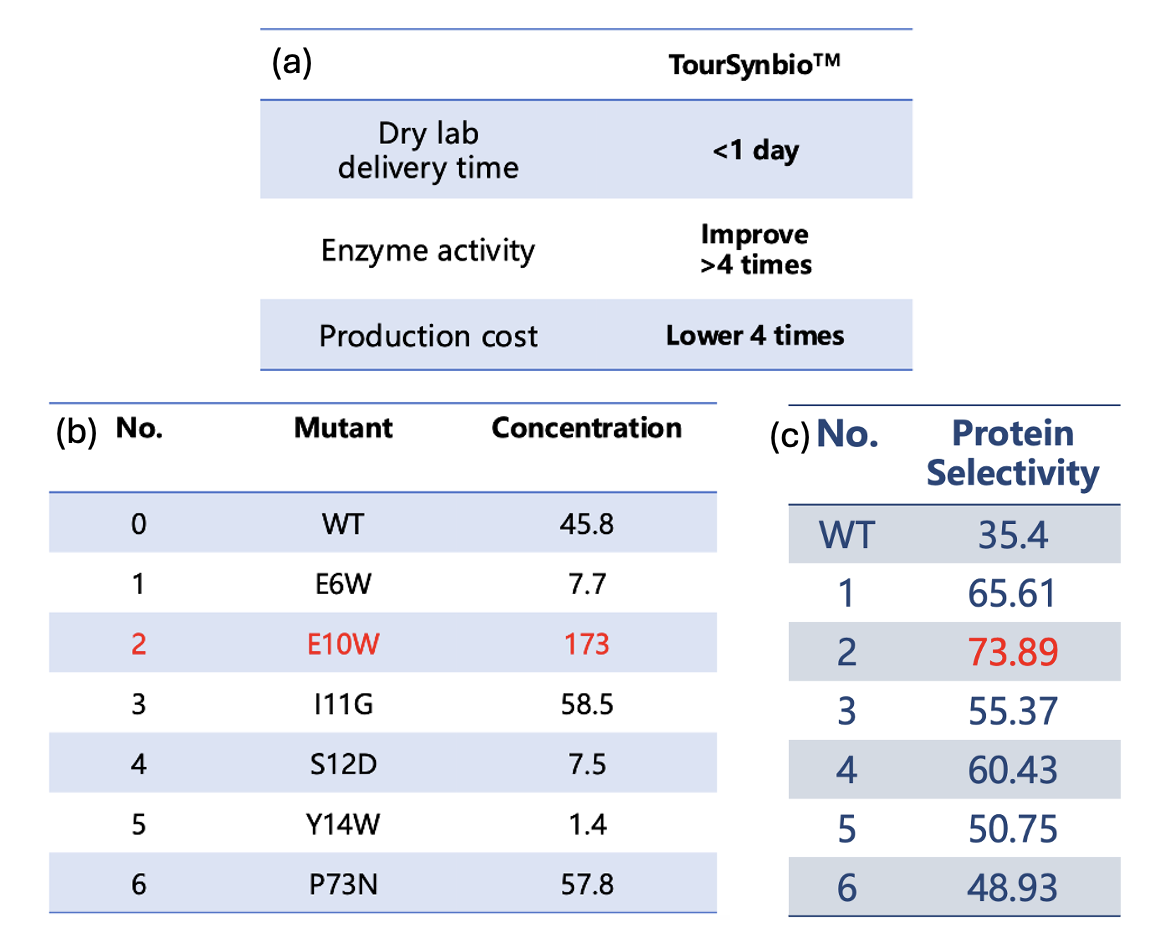}}
    \caption{Result on the vanilla key enzyme modification. 
    (a) Summary of TourSynbio-Agent's performance improvements in enzyme engineering: reduced dry lab delivery time to less than one day, enhanced enzyme activity by over four times, and lowered production costs by four times. 
    (b) Table showing various mutants and their concentrations, highlighting E10W in red due to its significantly higher concentration. 
    (c) Table displaying various protein mutants ranked by their selectivity, with the red mutant highlighted as the top performer.}
    \label{fig:result1}
\end{figure}

Vanilla, renowned for its distinctive aroma, is important in various industries, including cosmetics, food, and fragrance \cite{abbas2023aroma}. 
%
%
Natural vanilla extract, priced at 1,000 RMB/kg, is significantly more expensive than its synthetic counterpart (200 RMB/kg), yet preferred for high-end products due to superior quality. 
%
%
Protein engineering offers a promising solution to this challenge through the synthesis of vanilla from biological waste. 
This process relies on two-step enzymatic catalysis, with the Aldehyde Deformylating Oxygenase (ADO) protein playing an important role \cite{andre2013fusing}. 
The efficiency of ADO directly impacts the purity and yield of the produced vanilla, thereby influencing its market value.
Traditional rational design approaches for ADO optimization have shown limited success. 
%
%
%
Hence, we employed TourSynbio-7B and TouSynbio-Agent to address these challenges, aiming to reduce dry lab delivery time, improve enzyme turnover number, and decrease production costs, as depicted in Fig.~\ref{fig:result1}. 
%


\subsubsection{Steroid Compound Catalysis}
Steroid compounds are important for maintaining cell membrane fluidity and stability. 
%
%
%
Our case study focuses on improving the selectivity of the P450 enzyme, which catalyzes steroid compounds. 
By utilizing TourSynbio-7B and TourSynbio-Agent, we aim to identify potential modifications to the P450 enzyme that could enhance its catalytic selectivity, potentially leading to more efficient and precise steroid synthesis processes.
%


\section{Conclusion}
In this study, we introduced TourSynbio-7B, the first multi-modal large model specifically designed for protein engineering without relying on external protein encoders. 
TourSynbio-7B, leveraging the structural similarities between protein sequences and natural languages, demonstrated superior performance in protein understanding and engineering tasks, outperforming GPT-4 on the ProteinLMBench benchmark.
We further developed TourSynbio-Agent, an agent framework integrating various deep learning models and tools, offering a unified conversational interface for the efficient execution of complex protein engineering tasks. 
Our framework's capabilities were validated through wet lab case studies on enzyme modification and steroid compound catalysis, showcasing significant improvements in mutation accuracy, delivery time, and automation. 
Despite these advancements, TourSynbio-7B's performance in predicting highly complex protein structures could be further improved, indicating the need for specialized post-training on diverse structural datasets.
Future work will focus on integrating advanced structural prediction models to enhance the precision and applicability of TourSynbio-7B in more real-world protein engineering scenarios.


\section*{Acknowledgment}
\noindent We gratefully acknowledge Karamay Carbon Neutrality Network Technology Co., Ltd for providing the GPU resources used in this research.
%

\bibliographystyle{IEEEtran}
\bibliography{main.bib}

\begin{thebibliography}{10}
\providecommand{\url}[1]{#1}
\csname url@samestyle\endcsname
\providecommand{\newblock}{\relax}
\providecommand{\bibinfo}[2]{#2}
\providecommand{\BIBentrySTDinterwordspacing}{\spaceskip=0pt\relax}
\providecommand{\BIBentryALTinterwordstretchfactor}{4}
\providecommand{\BIBentryALTinterwordspacing}{\spaceskip=\fontdimen2\font plus
\BIBentryALTinterwordstretchfactor\fontdimen3\font minus \fontdimen4\font\relax}
\providecommand{\BIBforeignlanguage}[2]{{%
\expandafter\ifx\csname l@#1\endcsname\relax
\typeout{** WARNING: IEEEtran.bst: No hyphenation pattern has been}%
\typeout{** loaded for the language `#1'. Using the pattern for}%
\typeout{** the default language instead.}%
\else
\language=\csname l@#1\endcsname
\fi
#2}}
\providecommand{\BIBdecl}{\relax}
\BIBdecl

\bibitem{leisola2007protein}
M.~Leisola and O.~Turunen, ``Protein engineering: opportunities and challenges,'' \emph{Applied Microbiology and Biotechnology}, vol.~75, no.~6, pp. 1225--1232, 2007.

\bibitem{listov2024opportunities}
D.~Listov, C.~A. Goverde, B.~E. Correia, and S.~J. Fleishman, ``Opportunities and challenges in design and optimization of protein function,'' \emph{Nature Reviews Molecular Cell Biology}, pp. 1--15, 2024.

\bibitem{defresne2021protein}
M.~Defresne, S.~Barbe, and T.~Schiex, ``Protein design with deep learning,'' \emph{International Journal of Molecular Sciences}, vol.~22, no.~21, p. 11741, 2021.

\bibitem{dauparas2022robust}
J.~Dauparas \emph{et~al.}, ``Robust deep learning--based protein sequence design using proteinmpnn,'' \emph{Science}, vol. 378, no. 6615, pp. 49--56, 2022.

\bibitem{wang2018computational}
J.~Wang, H.~Cao, J.~Z. Zhang, and Y.~Qi, ``Computational protein design with deep learning neural networks,'' \emph{Scientific reports}, vol.~8, no.~1, pp. 1--9, 2018.

\bibitem{bairoch2000swiss}
A.~Bairoch and R.~Apweiler, ``The swiss-prot protein sequence database and its supplement trembl in 2000,'' \emph{Nucleic acids research}, vol.~28, no.~1, pp. 45--48, 2000.

\bibitem{heinzinger2019modeling}
M.~Heinzinger, A.~Elnaggar, Y.~Wang, C.~Dallago, D.~Nechaev, F.~Matthes, and B.~Rost, ``Modeling aspects of the language of life through transfer-learning protein sequences,'' \emph{BMC bioinformatics}, vol.~20, pp. 1--17, 2019.

\bibitem{rives2019biological}
\BIBentryALTinterwordspacing
A.~Rives \emph{et~al.}, ``Biological structure and function emerge from scaling unsupervised learning to 250 million protein sequences,'' \emph{PNAS}, 2019. [Online]. Available: \url{https://www.biorxiv.org/content/10.1101/622803v4}
\BIBentrySTDinterwordspacing

\bibitem{gpt4}
OpenAI \emph{et~al.}, ``{GPT-4 Technical Report},'' \emph{arxiv preprint arXiv:2303.08774}, 2024.

\bibitem{18}
\BIBentryALTinterwordspacing
M.~Blum \emph{et~al.}, ``{The InterPro protein families and domains database: 20 years on},'' \emph{Nucleic Acids Research}, vol.~49, no.~D1, pp. D344--D354, 11 2020. [Online]. Available: \url{https://doi.org/10.1093/nar/gkaa977}
\BIBentrySTDinterwordspacing

\bibitem{lv2024prollama}
L.~Lv, Z.~Lin, H.~Li, Y.~Liu, J.~Cui, C.~Y.-C. Chen, L.~Yuan, and Y.~Tian, ``Prollama: A protein large language model for multi-task protein language processing,'' \emph{arXiv preprint arXiv:2402.16445}, 2024.

\bibitem{wang2024protchatgpt}
C.~Wang, H.~Fan, R.~Quan, and Y.~Yang, ``Protchatgpt: Towards understanding proteins with large language models,'' \emph{arXiv preprint arXiv:2402.09649}, 2024.

\bibitem{xu_protst_2023}
\BIBentryALTinterwordspacing
M.~Xu, X.~Yuan, S.~Miret, and J.~Tang, ``{P}rot{ST}: Multi-modality learning of protein sequences and biomedical texts,'' in \emph{ICML}, vol. 202, 2023, pp. 38\,749--38\,767. [Online]. Available: \url{https://proceedings.mlr.press/v202/xu23t.html}
\BIBentrySTDinterwordspacing

\bibitem{zhou_protein_2023}
\BIBentryALTinterwordspacing
H.-Y. Zhou, Y.~Fu, Z.~Zhang, B.~Cheng, and Y.~Yu, ``Protein representation learning via knowledge enhanced primary structure reasoning,'' in \emph{ICLR}, 2023. [Online]. Available: \url{https://openreview.net/forum?id=VbCMhg7MRmj}
\BIBentrySTDinterwordspacing

\bibitem{in}
Z.~Cai \emph{et~al.}, ``Internlm2 technical report,'' \emph{arXiv preprint arXiv:2403.17297}, 2024.

\bibitem{data}
\BIBentryALTinterwordspacing
Y.~Shen, Z.~Chen, M.~Mamalakis, L.~He, H.~Xia, T.~Li, Y.~Su, J.~He, and Y.~G. Wang, ``A fine-tuning dataset and benchmark for large language models for protein understanding,'' 2024. [Online]. Available: \url{https://arxiv.org/abs/2406.05540}
\BIBentrySTDinterwordspacing

\bibitem{11}
\BIBentryALTinterwordspacing
S.~K. Burley \emph{et~al.}, ``{RCSB Protein Data Bank: powerful new tools for exploring 3D structures of biological macromolecules for basic and applied research and education in fundamental biology, biomedicine, biotechnology, bioengineering and energy sciences},'' \emph{Nucleic Acids Research}, vol.~49, no.~D1, pp. D437--D451, 11 2020. [Online]. Available: \url{https://doi.org/10.1093/nar/gkaa1038}
\BIBentrySTDinterwordspacing

\bibitem{6}
\BIBentryALTinterwordspacing
P.~Liang \emph{et~al.}, ``{Holistic Evaluation of Language Models},'' \emph{Transactions on Machine Learning Research}, 2023, featured Certification, Expert Certification. [Online]. Available: \url{https://openreview.net/forum?id=iO4LZibEqW}
\BIBentrySTDinterwordspacing

\bibitem{13}
L.~Kinch, R.~Schaeffer, A.~Kryshtafovych, and N.~Grishin, ``\BIBforeignlanguage{English (US)}{{Target classification in the 14th round of the critical assessment of protein structure prediction (CASP14)}},'' \emph{\BIBforeignlanguage{English (US)}{Proteins: Structure, Function and Bioinformatics}}, vol.~89, no.~12, pp. 1618--1632, Dec. 2021.

\bibitem{4}
\BIBentryALTinterwordspacing
D.~Hendrycks \emph{et~al.}, ``Measuring massive multitask language understanding,'' in \emph{ICLR}, 2021. [Online]. Available: \url{https://openreview.net/forum?id=d7KBjmI3GmQ}
\BIBentrySTDinterwordspacing

\bibitem{kapoor2024ai}
S.~Kapoor, B.~Stroebl, Z.~S. Siegel, N.~Nadgir, and A.~Narayanan, ``Ai agents that matter,'' \emph{arXiv preprint arXiv:2407.01502}, 2024.

\bibitem{d3}
\BIBentryALTinterwordspacing
Z.~Lin, H.~Akin \emph{et~al.}, ``Language models of protein sequences at the scale of evolution enable accurate structure prediction,'' \emph{bioRxiv}, 2022. [Online]. Available: \url{https://www.biorxiv.org/content/early/2022/07/21/2022.07.20.500902}
\BIBentrySTDinterwordspacing

\bibitem{hollingsworth2018molecular}
S.~A. Hollingsworth and R.~O. Dror, ``Molecular dynamics simulation for all,'' \emph{Neuron}, vol.~99, no.~6, pp. 1129--1143, 2018.

\bibitem{s7}
\BIBentryALTinterwordspacing
J.~Su, C.~Han \emph{et~al.}, ``Saprot: Protein language modeling with structure-aware vocabulary,'' \emph{bioRxiv}, 2023. [Online]. Available: \url{https://www.biorxiv.org/content/early/2023/10/02/2023.10.01.560349.1}
\BIBentrySTDinterwordspacing

\bibitem{s6}
\BIBentryALTinterwordspacing
J.~Meier, R.~Rao \emph{et~al.}, ``Language models enable zero-shot prediction of the effects of mutations on protein function,'' \emph{bioRxiv}, 2021. [Online]. Available: \url{https://www.biorxiv.org/content/early/2021/07/10/2021.07.09.450648}
\BIBentrySTDinterwordspacing

\bibitem{s8}
\BIBentryALTinterwordspacing
J.~Smirnovien{\.e}, L.~Baranauskien{\.e}, A.~Zubrien{\.e}, and D.~Matulis, ``A standard operating procedure for an enzymatic activity inhibition assay,'' \emph{European Biophysics Journal}, vol.~50, no.~3, pp. 345--352, 2021. [Online]. Available: \url{https://doi.org/10.1007/s00249-021-01530-8}
\BIBentrySTDinterwordspacing

\bibitem{outeiral2024codon}
C.~Outeiral and C.~M. Deane, ``Codon language embeddings provide strong signals for use in protein engineering,'' \emph{Nature Machine Intelligence}, vol.~6, no.~2, pp. 170--179, 2024.

\bibitem{prev}
\BIBentryALTinterwordspacing
Y.~Shen, O.~Lv, H.~Zhu, and Y.~G. Wang, ``Proteinengine: Empower llm with domain knowledge for protein engineering,'' 2024. [Online]. Available: \url{https://arxiv.org/abs/2405.06658}
\BIBentrySTDinterwordspacing

\bibitem{s5}
\BIBentryALTinterwordspacing
C.~Hsu, R.~Verkuil, J.~Liu, Z.~Lin, B.~Hie, T.~Sercu, A.~Lerer, and A.~Rives, ``Learning inverse folding from millions of predicted structures,'' \emph{bioRxiv}, 2022. [Online]. Available: \url{https://www.biorxiv.org/content/early/2022/09/06/2022.04.10.487779}
\BIBentrySTDinterwordspacing

\bibitem{yi2024graph}
K.~Yi, B.~Zhou, Y.~Shen, P.~Li{\`o}, and Y.~Wang, ``Graph denoising diffusion for inverse protein folding,'' \emph{Advances in Neural Information Processing Systems}, vol.~36, 2024.

\bibitem{s4}
\BIBentryALTinterwordspacing
M.~H. Høie \emph{et~al.}, ``Antifold: Improved antibody structure-based design using inverse folding,'' 2024. [Online]. Available: \url{https://arxiv.org/abs/2405.03370}
\BIBentrySTDinterwordspacing

\bibitem{s3}
\BIBentryALTinterwordspacing
J.~B. Ingraham \emph{et~al.}, ``Illuminating protein space with a programmable generative model,'' \emph{Nature}, vol. 623, no. 7989, pp. 1070--1078, 2023. [Online]. Available: \url{https://doi.org/10.1038/s41586-023-06728-8}
\BIBentrySTDinterwordspacing

\bibitem{s1}
J.~Dauparas \emph{et~al.}, ``Atomic context-conditioned protein sequence design using ligandmpnn,'' \emph{Biorxiv}, pp. 2023--12, 2023.

\bibitem{lagent2023}
L.~D. Team, ``{Lagent: InternLM} a lightweight open-source framework that allows users to efficiently build large language model(llm)-based agents,'' \url{https://github.com/InternLM/lagent}, 2023.

\bibitem{f}
E.~Almazrouei, H.~Alobeidli, A.~Alshamsi, A.~Cappelli, R.~Cojocaru, M.~Debbah, Étienne Goffinet, D.~Hesslow, J.~Launay, Q.~Malartic, D.~Mazzotta, B.~Noune, B.~Pannier, and G.~Penedo, ``The {Falcon Series of Open Language Models},'' \emph{arXiv preprint arXiv:2311.16867}, 2023.

\bibitem{q}
J.~Bai \emph{et~al.}, ``Qwen technical report,'' \emph{arXiv preprint arXiv:2309.16609}, 2023.

\bibitem{moon}
D.~J. Zhang \emph{et~al.}, ``Moonshot: Towards controllable video generation and editing with multimodal conditions,'' \emph{arXiv preprint arXiv:2401.01827}, 2024.

\bibitem{m}
A.~Q. Jiang \emph{et~al.}, ``{Mistral 7B},'' \emph{arXiv preprint arXiv:2310.06825}, 2023.

\bibitem{b}
\BIBentryALTinterwordspacing
Baichuan, ``Baichuan 2: Open large-scale language models,'' \emph{arXiv preprint arXiv:2309.10305}, 2023. [Online]. Available: \url{https://arxiv.org/abs/2309.10305}
\BIBentrySTDinterwordspacing

\bibitem{ll}
H.~Touvron \emph{et~al.}, ``{Llama 2: Open Foundation and Fine-Tuned Chat Models},'' \emph{arXiv preprint arXiv:2307.09288}, 2023.

\bibitem{ch}
Z.~Du, Y.~Qian, X.~Liu, M.~Ding, J.~Qiu, Z.~Yang, and J.~Tang, ``{GLM: General Language Model Pretraining with Autoregressive Blank Infilling},'' in \emph{ACL (Volume 1: Long Papers)}, 2022, pp. 320--335.

\bibitem{yi}
01.AI, A.~Y. : \emph{et~al.}, ``{Yi: Open Foundation Models by 01.AI},'' \emph{arXiv preprint arXiv:2403.04652}, 2024.

\bibitem{abbas2023aroma}
F.~Abbas \emph{et~al.}, ``Aroma components in horticultural crops: chemical diversity and usage of metabolic engineering for industrial applications,'' \emph{Plants}, vol.~12, no.~9, p. 1748, 2023.

\bibitem{andre2013fusing}
C.~Andre \emph{et~al.}, ``Fusing catalase to an alkane-producing enzyme maintains enzymatic activity by converting the inhibitory byproduct h2o2 to the cosubstrate o2,'' \emph{Proceedings of the National Academy of Sciences}, vol. 110, no.~8, pp. 3191--3196, 2013.

\end{thebibliography}

\appendix
\section{Case study}
\subsection{Case 1: Enhancing the Activity of Key Enzymes for Vanillin Production for Our Clients}
Vanillin has widespread market demand, serving as a raw material extensively used in food, fragrances, and cosmetics industries, with a global market sales of around 440 million dollars. Chemically synthesized vanillin is priced below 200 yuan/kg, while natural vanillin is estimated at 20,000 yuan/kg. The cost of biosynthesized vanillin is around 3,000 yuan/kg, with quality comparable to naturally extracted vanillin, but higher priced than chemical synthesis. If the cost of 3,000 yuan/kg can be further reduced, it will significantly enhance the product's market competitiveness.

\begin{figure}[h!]
    \centering
    \centerline{\includegraphics[width=\linewidth]{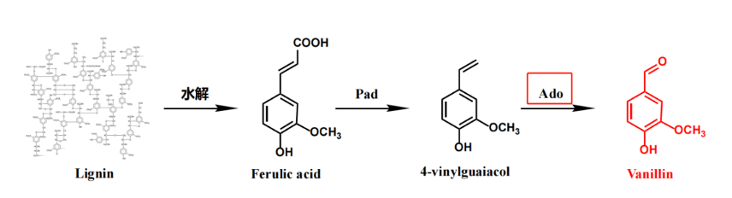}}
    \caption{Lignin is converted to vanillin through a two-step biocatalytic process, where the activity of the Ado enzyme in the second step is crucial for enhancing yield.}
    \label{fig:case1}
\end{figure}
 
We biosynthesize vanillin from biological waste through a two-step enzymatic catalysis process. The key enzyme Ado determines the purity, yield and other properties of vanillin, thereby affecting the product's market value.  

In traditional methods, rational design is a common approach for enzyme modification, analyzing the protein's 3D structure, active site pockets, active site residues, amino acid hydrophobicity/charge properties etc. to propose potential modification sites and alternative amino acid types. However, this method requires extensive theoretical and application knowledge in biology, is highly competitive, and the number of alternative options is typically limited by the knowledge reserve. High-throughput multi-site full amino acid type mutagenesis experiments can increase screening possibilities but at extremely high costs.
After employing rational design, LumbyBio identified 6 potential activity-enhancing sites on the Ado protein and generated 114 mutation schemes across all amino acid types. However, experiments on the 4-VG substrate showed none of these mutations could improve enzyme activity, prompting LumbyBio to collaborate with us.
TourSynbio designed single-site modification schemes for the 595 amino acid Ado protein using the enzyme modification model in ProteinEngine, recommending 19 mutant amino acid sequences. 12 were selected for wet-lab experiments to verify enzyme activity. Leveraging the 114 wet experiment data from LumbyBio to fine-tune the downstream modules of TM, new amino acid sequences predicted to enhance protein activity were recommended.
Ultimately, we successfully delivered the product, with the recommended E10W mutation exhibiting 175 expression level compared to 45.8 for the wild-type, a 4-fold increase significantly reducing production costs.

\subsection{Case 2: Enhancing Steroid Compound Selectivity for Our Clients}

\begin{figure}[h!]
    \centering
    \centerline{\includegraphics[width=\linewidth]{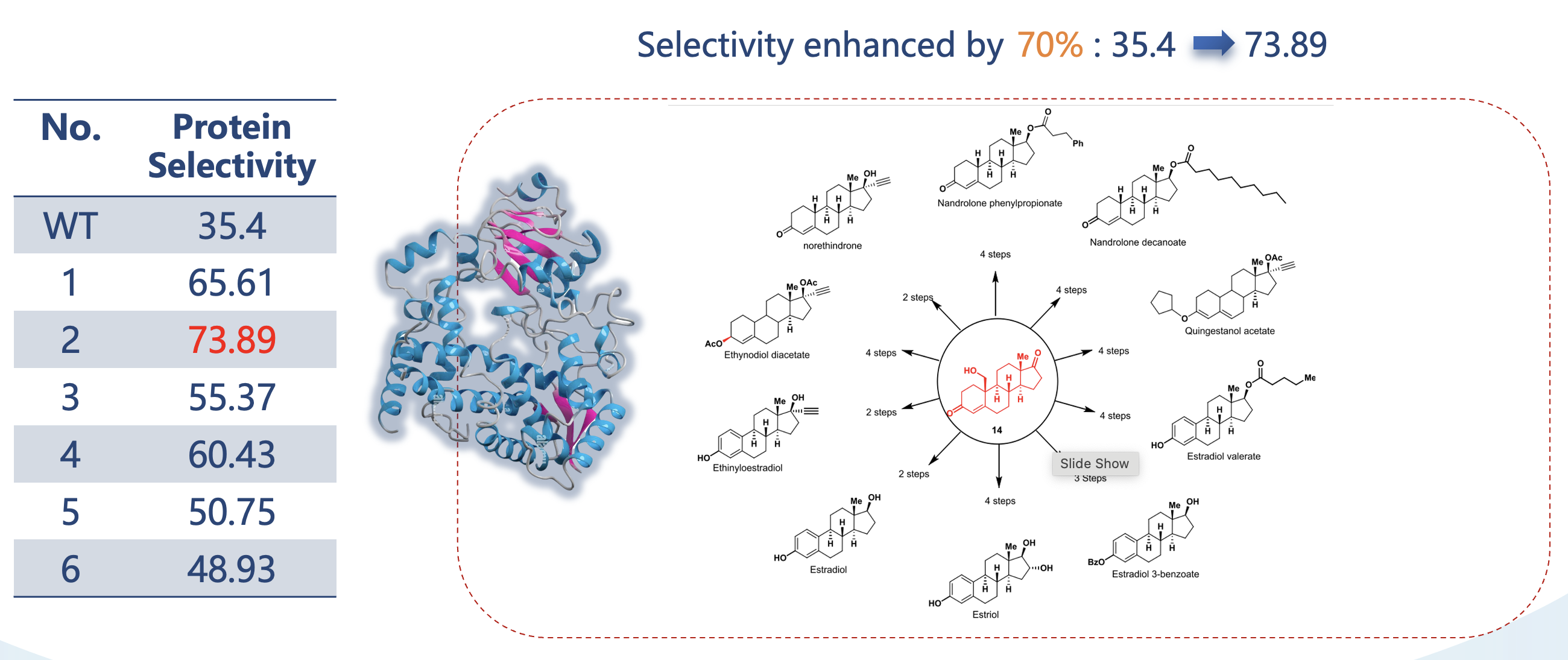}}
    \caption{The goal is to modify the P450 protein, which catalyzes the 19-hydroxylation of steroid compounds, to increase its selectivity by 70\% for the effective product, a crucial step for scaling up production efficiency.}
    \label{fig:case2}
\end{figure}

Steroid compounds are a class of biologically active molecules widely found in nature. They play a key role in the human body, including maintaining the fluidity and stability of cell membranes, and participating in various physiological processes as hormones. In addition, steroid compounds are used as non-hormonal drugs for the treatment of cardiovascular and cerebrovascular diseases, among others. There are more than 400 steroid drugs approved for marketing in the global market, with the steroid drug market size reaching 100 billion US dollars, covering a variety of therapeutic areas such as anti-tumor, anti-convulsant, anti-allergy, anti-fungal, anti-inflammatory, immunosuppressive, anti-androgen, and diuretic.
 
In this case, we focus on the P450 protein that catalyzes the 19-hydroxylation of steroid compounds. P450 proteins are an important class of enzymes involved in various biosynthetic and metabolic processes, including the synthesis of steroid compounds. When catalyzing steroid compounds, P450 proteins produce four different products, but only one of them is effective. The goal of the modification is to increase the proportion of this effective product, that is, to improve the selectivity of the enzyme catalysis. To meet the transition from the laboratory to production scale, the client set a target: to increase selectivity by 70\%. This enhancement is key to achieving production efficiency.

\subsubsection{Modification Strategy}
\paragraph{First round of modification} Using the TourSynbio large model, we recommended 200 single-site mutations within two weeks. These mutations were then tested in wet experiments over three weeks to evaluate their impact on selectivity.
\paragraph{Second round of modification} Using the 200 wet experiment data, we fine-tuned the TourSynbio business model and ultimately recommended 10 mutations with no more than five sites.

\subsubsection{Modification Results}
\paragraph{Correlation assessment} The correlation between the recommended mutation results and experimental data reached 0.7, indicating that our model has high predictive accuracy.
\paragraph{Selectivity improvement} Among the 10 recommended results, selectivity was increased by up to 70\%, and the enzyme activity was also slightly improved.
\paragraph{Production readiness} These results basically met the requirements for pushing the P450 protein from the laboratory to production, paving the way for efficient production of steroid compounds.

Through AI-assisted enzyme modification, we successfully improved the selectivity of P450 protein catalysis for the 19-hydroxylation of steroid compounds, providing a new solution to produce steroid drugs. This case demonstrates the potential of AI technology in the field of bioengineering, especially in improving the efficiency and selectivity of biocatalysts.

\subsection{Case 3: Assisting customers in enhancing the catalytic conversion rate of enzymes}

Alcohol compounds, especially hormone drugs, hold a pivotal position in the global pharmaceutical market. Progesterone, testosterone, estradiol, cortisol, aldosterone, and synthetic progestogens are not only crucial to human health but also have significant economic benefits. These compounds play a key role in treating a variety of diseases, ranging from reproductive health to metabolic disorders, and from inflammation to immune regulation. With the aging of the global population and the increase in lifestyle-related diseases, the demand for these drugs is expected to continue to grow.

\begin{figure}[h!]
    \centering
    \centerline{\includegraphics[width=\linewidth]{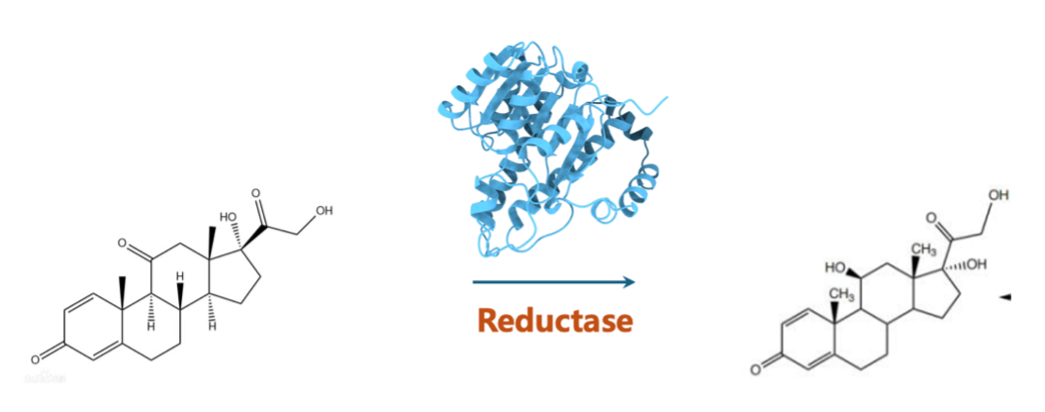}}
    \caption{Reductase catalysis of alcohol compounds.}
    \label{fig:case3}
\end{figure}

Improving the catalytic conversion rate of reductase is the core goal of this case. An increase in conversion rate means that more substrate is converted into useful products within the same reaction time, thereby improving production efficiency and reducing costs.

Using the wild-type sequence and 29 single-point mutation data provided by the customer, the business model of TuShen was fine-tuned to recommend 10 single-point mutation enzyme variants. The recommended mutations were tested for 4 weeks in wet experiments to evaluate their conversion rates.

\subsubsection{Modification results}
\paragraph{Correlation assessment} The correlation between the recommended results and wet experiment data reached 0.7, showing a high consistency between model predictions and experimental results.
\paragraph{Conversion rate improvement} Among the 10 recommended results, the conversion rate was improved by up to 3.7 times, significantly enhancing catalytic efficiency.

Through AI-assisted enzyme modification, we have improved the catalytic conversion rate of reductase, providing a new solution to produce alcohol compounds. The improved conversion rate directly reduces production costs and increases the utilization rate of raw materials. It has strengthened the competitiveness of the customer's company in the hormone drug market, bringing exclusive market opportunities and high profits to companies developing and producing these drugs.

\end{document}